\documentclass[aps,prl,twocolumn,superscriptaddress]{revtex4}
\usepackage{graphicx}
\usepackage{times}
\usepackage{amsmath}
\usepackage{textcomp}
\usepackage{epstopdf}
\usepackage{float}

\begin{document}

\title{Spin-orbit coupling and operation of multi-valley spin qubits}

\author{M. Veldhorst}
\email{M.Veldhorst@unsw.edu.au}
\affiliation{Centre for Quantum Computation and Communication
Technology, School of Electrical Engineering and Telecommunications,
The University of New South Wales, Sydney, NSW 2052, Australia}
\author{R. Ruskov}
\affiliation{Laboratory for Physical Sciences, 8050 Greenmead Dr.,
College Park, MD 20740, USA.}
\author{C.H. Yang}
\affiliation{Centre for Quantum Computation and Communication
Technology, School of Electrical Engineering and Telecommunications,
The University of New South Wales, Sydney, NSW 2052, Australia}
\author{J.C.C. Hwang}
\affiliation{Centre for Quantum Computation and Communication
Technology, School of Electrical Engineering and Telecommunications,
The University of New South Wales, Sydney, NSW 2052, Australia}
\author{F.E. Hudson}
\affiliation{Centre for Quantum Computation and Communication
Technology, School of Electrical Engineering and Telecommunications,
The University of New South Wales, Sydney, NSW 2052, Australia}
\author{M.E. Flatt$\acute{\textrm e}$}
\affiliation{Department of Physics and Astronomy, University of
Iowa, Iowa City, Iowa 52242.}
\author{C. Tahan}
\affiliation{Laboratory for Physical Sciences, 8050 Greenmead Dr.,
College Park, MD 20740, USA.}
\author{K.M. Itoh}
\affiliation{School of Fundamental Science and Technology, Keio
University, 3-14-1 Hiyoshi, Kohoku-ku, Yokohama 223-8522, Japan.}
\author{A. Morello}
\affiliation{Centre for Quantum Computation and Communication
Technology, School of Electrical Engineering and Telecommunications,
The University of New South Wales, Sydney, NSW 2052, Australia}
\author{A.S. Dzurak}
\email{A.Dzurak@unsw.edu.au}
\affiliation{Centre for Quantum Computation and Communication
Technology, School of Electrical Engineering and Telecommunications,
The University of New South Wales, Sydney, NSW 2052, Australia}

\date{\today}

\begin{abstract}
Spin qubits composed of either one or three electrons are realized in a quantum dot formed at a
Si/SiO$_2$ interface in isotopically enriched silicon. Using pulsed electron spin resonance,
we perform coherent control of both types of qubits, addressing
them via an electric field dependent $g$-factor. We perform randomized benchmarking
and find that both qubits can be operated with high fidelity. Surprisingly, we find that the $g$-factors
of the one-electron and three-electron qubits have an approximately linear but opposite dependence as a function of
the applied dc electric field. We develop a theory to explain
this $g$-factor behavior based on the
spin-valley coupling that results from the sharp interface.
The outer ``shell'' electron in the
three-electron qubit exists in the higher of the two available conduction-band valley states, in contrast
with the one-electron case, where the electron is in the lower valley.
We formulate a modified effective mass theory and
propose that inter-valley spin-flip tunneling dominates over intra-valley spin-flips
in this system, leading to a direct correlation between the spin-orbit coupling parameters and
the $g$-factors in the two valleys.
In addition to offering all-electrical tuning for single-qubit gates,
the $g$-factor physics revealed here for one-electron and three-electron qubits offers potential
opportunities for new qubit control approaches.
\end{abstract}

\maketitle

Silicon is known among the semiconductors to have small spin-orbit
coupling (SOC), a beneficial fact for silicon quantum computing, since charge
noise is largely decoupled from information stored in the spin
\cite{Zwanenburg2013}. Furthermore, silicon can be
isotopically enriched and chemically purified to $^{28}$Si, thereby removing
nuclear spin background fluctuations and so silicon is often referred to
as a semiconductor vacuum
\cite{Itoh2014}. These two facts have motivated intense
research on silicon qubits,
leading to recent realizations of
single-qubit \cite{Maune2012, Pla2012, Kim2014, Kawakami2014,
Veldhorst2014} and two-qubit \cite{Veldhorst20142} logic gates.
Despite the small SOC in
isotopically purified silicon quantum dots, the small tunability of
the $g$-factor via gate-controlled electric fields allows one to electrostatically turn
on and off the spin rotations that constitute single-qubit gates
\cite{Veldhorst2014, Veldhorst20142, Laucht2015}, thereby providing an important tool for quantum
computation. 

\begin{figure}[!h] 
    \centering
        \includegraphics[width=0.38\textwidth]{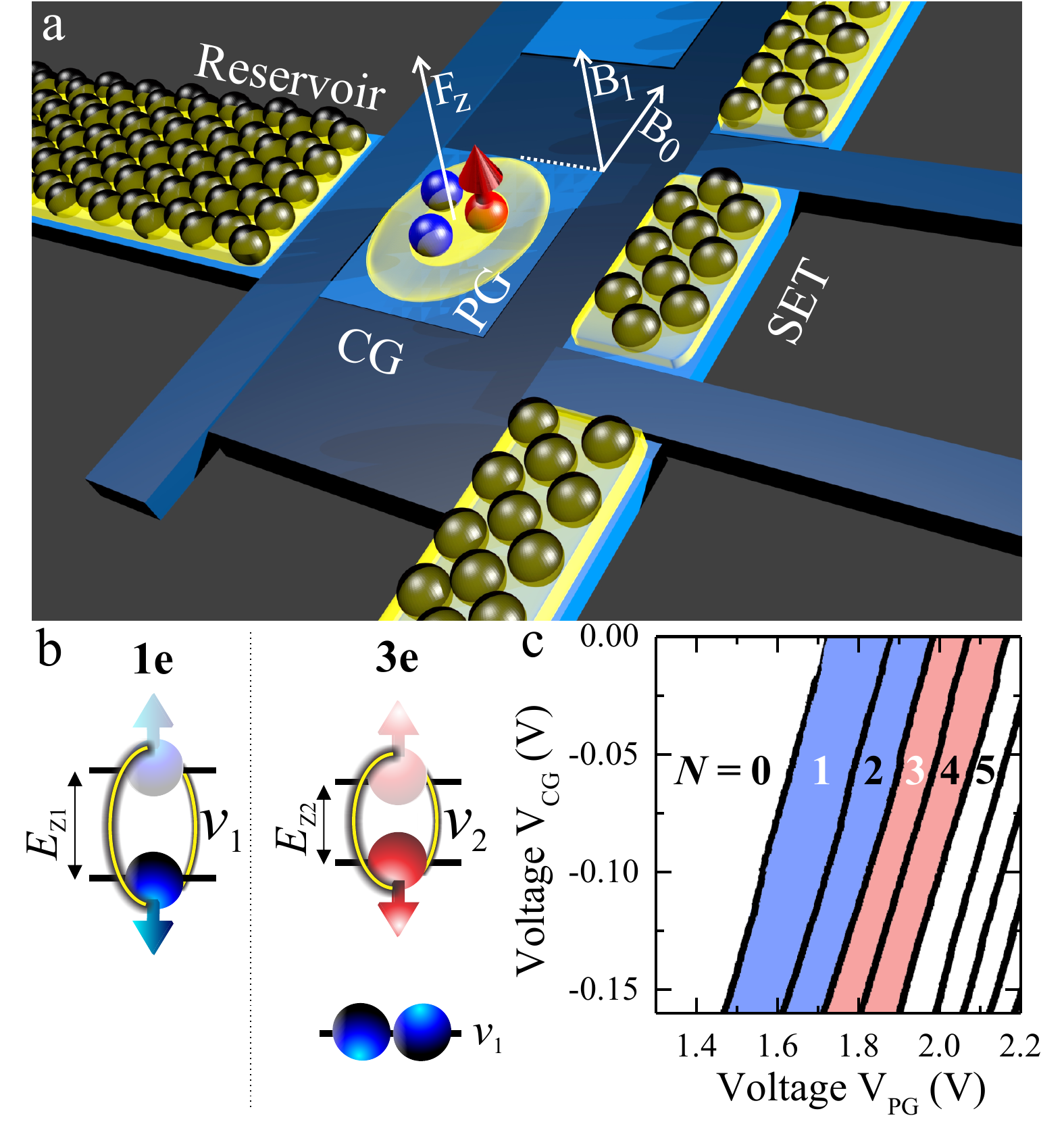}
        \caption{ (a) Schematic representation of the quantum dot system.
        The quantum dot is defined using the confinement gate $CG$ and plunger gate $PG$ and the yellow shading represents
        the regions where electrons are accumulated, with $F_z$ 
        the perpendicular electric field direction.
        ESR control is via a dc magnetic field
                $B_0 = 1.4\,{\rm T}$ (in the Si $[110]$ in-plane direction) and an ac magnetic field $B_1$.
        (b) The quantum dot qubit can be operated using the spin states of one electron, or
        using three electrons, where two electrons (blue) occupy the lowest energy valley state
        and the third electron (red) is in the higher energy valley state.
        (c) Charge stability diagram showing the electron occupancy $N$ in the quantum dot, measured with a nearby SET.}
        \label{fig:1}
\end{figure}

The low-energy subspace in silicon quantum dot (QD) systems is governed
by two spin-degenerate valley states. When these valley states are
quasi-degenerate, qubit operation becomes complex
\cite{Kawakami2014}, and the coupling of qubits is even more challenging
\cite{Culcer2010}. However, the valley states can be separated using
a vertical electric field and the sharp potential of an interface \cite{Yang2013, Veldhorst2014},
and their energy separation can be electrically controlled over several
hundreds of $\mu{\rm eV}$.   
While one-electron spin qubits  
are naturally operated in the lowest valley state
\cite{Kawakami2014, Veldhorst2014},
it is intriguing to consider the potential performance of qubits operated in the higher valley state,
which is also known to have a long spin lifetime if orbital relaxation can be suppressed \cite{Yang2013}.
When spatial confinement in the QD  
is strong, the orbital excited states are
lifted high in energy and qubit operation in the upper valley state is
possible by populating three electrons in the quantum dot. In this mode two
electrons form a singlet in the lower valley state and the third
electron is operated in the upper valley
(see Fig.\ref{fig:1}a and \ref{fig:1}b).
It has been suggested that such multi-electron qubits could enhance the gate
fidelity, due to partial screening of electrical noise
\cite{Barnes2011}.

In this letter, we demonstrate high-fidelity operation of one- and three-electron spin qubits,
operated in the lower and upper
valley, respectively. Using pulsed electron-spin-resonance (ESR) we
map out with high precision the qubit frequency as a function of the applied
perpendicular electric field, controlled with electrostatic gates.
We experimentally demonstrate and theoretically explain how
inter-valley spin-orbit coupling at the  
Si/SiO$_2$ interface
results in an opposite dependence of the  
$g$-factor
for the two valleys, that is correlated with the
corresponding  dependence of the 2D Rashba and Dresselhaus
spin-orbit coupling in each valley.
We also present randomized benchmarking on both qubit systems,
showing   
they are capable of fidelities above 99\%, approaching the
surface
code thresholds for fault-tolerant quantum computing
\cite{Fowler2012}.

\begin{figure}
    \centering
        \includegraphics[width=0.45\textwidth]{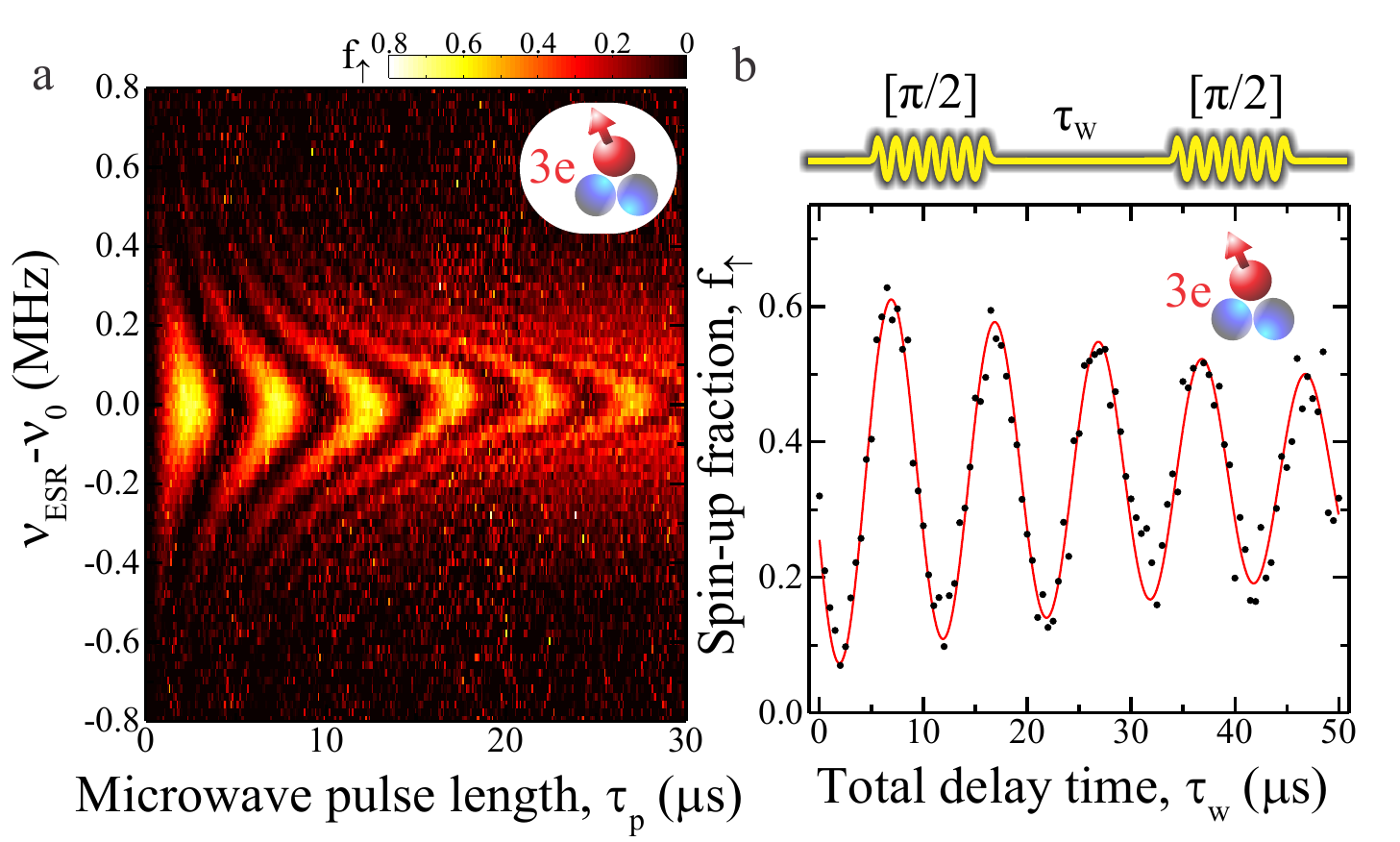}
        \caption{Demonstration of qubit control of the three-electron qubit.
        (a) 2D colour map showing Rabi control of the spin-up fraction $f_{\uparrow}$,
        by varying the microwave pulse length and the microwave driving frequency $\nu_{ESR}$.
                We have subtracted a reference frequency $\nu_0 = 39.045$ GHz (corresponding to $g = 1.9908$) for clarity.
        (b) Ramsey fringes, obtained by varying the waiting time in between two ESR $\pi/2$ pulses.
        The decay in the spin-up fraction $f_{\uparrow}$ corresponds to $T_2^*$ = 70 $\mu$s.
        The confinement gate voltage
        is $V_C$ = -0.2 V.}
        \label{fig:2}
\end{figure}

The 
QD structure is fabricated on an
epitaxially grown, isotopically purified $^{28}$Si epilayer with a
residual concentration of $^{29}$Si at 800 ppm \cite{Itoh2014}. The
aluminum gates are fabricated with electron-beam lithography using a
multi-level gate stack silicon metal-oxide-semiconductor (SiMOS) technology \cite{Angus2007},
see Fig. \ref{fig:1}.
The charge stability diagram of the quantum dot is shown in Fig. \ref{fig:1}c.
We use ESR to control the one-electron qubit \cite{Veldhorst2014} and the three-electron qubit,
see Fig. \ref{fig:2}. From a Ramsey sequence we find a long dephasing time $T_2^* = 70\, \mu {\rm s}$,
which is slightly less than we have previously measured for the one-electron qubit, which had
$T_2^* = 120\, \mu {\rm s}$ \cite{Veldhorst2014}.

We have demonstrated electric field control over the resonance frequency $\nu_{\rm ESR}$ of the one-electron qubit
\cite{Veldhorst2014},
 showing tunability over several MHz that appears linear in electric field, corresponding to more than 3000 times the
$2.4\,{\rm kHz}$  ESR line-width. We find that spin-valley mixing of the QD eigenstates
due to interface roughness \cite{Yang2013} would predict a modification of the electron
$g$-factor that is two orders of magnitude smaller than is found experimentally,
together with a non-linear dependence close to the anticrossing point of the
spin-valley states that we do not observe. Here we propose and analyze a model where the $g$-factor modification proceeds
via inter-valley spin-flip tunneling, mediated by the strong $z$-confinement at the interface.
The $\rm Si/SiO_2$ (001) interface of silicon MOS quantum dots can be
described with a Hamiltonian that consists of a bulk term
${\cal H}_0$ and an interface term ${\cal H}_{if}$.
The reduction of the bulk Si crystal symmetry at the interface,
in the presence of strong perpendicular electron confinement
induced by an applied electric field $F_z$, lifts the six-fold
valley degeneracy, leaving two low-lying  
$\Delta$-valleys at $\pm k_{0z}$.
These are then mixed via enhanced inter-valley tunneling
due to the strong $z$-confinement at the interface \cite{Ohkawa1978, Ando1982}.
The consequent effective
two-valley Hamiltonian acts on the four-component vector
$(\Phi_{\hat{z},\uparrow}(r),\Phi_{\hat{z},\downarrow}(r),
\Phi_{-\hat{z},\uparrow}(r),\Phi_{-\hat{z},\downarrow}(r) )^T$
$\equiv\Phi(r)$,
where the bulk part (spin and valley degenerate)
is given by
\begin{equation}
{\cal H}_0 =  \left[ \sum_{j=x,y,z} \frac{\hbar^2 \hat{k}_j^2}{2 m_j}
+ U_{x,y} + U_z \right]\times \hat{\bm{I}}_4 \label{bulk_H}
\end{equation}
with the quasi-momentum operators $\hat{k}_j \equiv -i \partial_j$;
and $U_{x,y} = \frac{m_t}{2} \omega_0^2( x^2 + y^2)$ 
and $U_z$ = $|e|F_z z$ are the in-plane and perpendicular confinement electron potentials, respectively.
Here $m_l$, $m_t$ are the
Si effective $\Delta$-valley electron masses,
$|e|$ is the electron charge,
and $\hbar$ is the reduced Planck constant.
Taking into account the large band offset of $\rm Si/SiO_2$,
the interface term is   
\begin{eqnarray}
&&{\cal H}_{\rm if} = -\frac{\hbar^2}{2 R m_l} \delta(z-z_0)
- i \frac{\hbar^2}{2 m_l} \delta(z-z_0) \hat{k}_z
\nonumber \\
&& \qquad\qquad\qquad  {} + \delta(z-z_0) \hat{V}_{\rm if}(\bm{k})
\label{interface_H},
\end{eqnarray}
where $R$ is
a parameter with dimension of length,
characterizing an abrupt interface \cite{Volkov1979,Vasko1979}, and
$|R| \ll l_z \ll l_D$;
here $l_z = (\hbar^2/2m_l |e| F_z)^{1/3}$ and $l_D = (\hbar/m_t \omega_0)^{1/2}$
are the perpendicular and in-plane confinement lengths
(assuming much stronger $\hat{z}$-confinement).
For $R \approx 0$ the interface Hamiltonian (Eq. \ref{interface_H})
corresponds to the standard infinite boundary condition (BC)
$\Phi(z)\mid_{z=z_0} = 0$, while for finite $R$  it generates
spin and valley mixing at the interface, $z \gtrsim z_0$.
Following the symmetry reasoning of
Refs. \cite{Golub2004,Nestoklon2006,Nestoklon2008} the spin-valley
mixing interface matrix $\hat{V}_{\rm if}(\bm{k})$ can be expressed
via the $C_{2v}$ invariants
$H_R(\bm{k}) = \sigma_x k_y - \sigma_y k_x$,
$H_D(\bm{k}) = \sigma_x k_x - \sigma_y k_y$, resulting in
\begin{equation}
\hat{V}_{\rm if}(\bm{k}) = \left[
\begin{array}{cc}
 A(\bm{k}) & V \hat{\bm{I}}_2  + B(\bm{k})\\
 V^* \hat{\bm{I}}_2 + B^{\dagger}(\bm{k})& A(\bm{k})
\end{array}
\right] \label{spin-valley-mixing} .
\end{equation}
In  Eq.~(\ref{spin-valley-mixing}) the $2\times 2$  block-diagonal
element $A(\bm{k})\equiv s_D\, H_D(\bm{k}) + s_R\, H_R(\bm{k})$
corresponds to \textit{intra-valley} spin-flipping transitions, while the
off-diagonal elements $V \hat{\bm{I}}_2$,
$B(\bm{k})\equiv \chi_D\, H_D(\bm{k}) + \chi_R\, H_R(\bm{k})$
are related to \textit{inter-valley}
tunneling (in momentum space) with no spin-flipping or with a
spin-flipping process.
Since experimentally the valley splitting ($\sim |V|$) is generally large
with respect to the spin-flipping terms,  
it makes sense to diagonalize with respect to the leading $V$-matrix element
and via a unitary transformation we find
\begin{equation}
\hat{V}^{\rm U}_{\rm if}(\bm{k}) = \left[
\begin{array}{cc}
|V|+ A + \frac{1}{2}B_{\rm d} & \frac{1}{2}B_{\rm off}\\
 {\rm h.c.} & -|V|+ A - \frac{1}{2}B_{\rm d}
\end{array}
\right] \label{spin-valley-mixing-unitary-transformed} .
\end{equation}
This matrix is approximately diagonal in the valley
basis $|v_1\rangle$, $|v_2\rangle$,
with a calculated valley splitting energy
$E_{\rm VS} = 2|V| R^2 |\varphi^{\prime}(0)|^2 = 2|V| R^2 l_z^{-3} \propto F_z$.
We neglect the off-diagonal
contribution $B_{\rm off} \equiv B - B^{\dagger} e^{2i\phi_V}$
in Eq.(\ref{spin-valley-mixing-unitary-transformed}), since it is
suppressed as $\sim 1/E_{\rm VS}$ and $E_{\rm VS}$ is typically several hundreds of $\mu{\rm eV}$ in MOS quantum dots
\cite{Yang2013, Veldhorst2014}.
Thus, in the valley
subspaces $|v_1\rangle$, $|v_2\rangle$,
one can consider two independent boundary conditions as in Eq.(\ref{interface_H}),
with spin-flipping interface matrices
$\hat{V}_{v1,v2} = A \mp \frac{1}{2}B_{\rm d} \equiv
A \mp \frac{1}{2}(B e^{-i\phi_V}+ B^{\dagger} e^{i\phi_V})$,
in which the inter-valley spin-flip tunneling element
changes sign between $v1$ and $v2$.

The effective 2D spin-orbit Hamiltonians
(proportional to the Rashba and Dresselhaus forms, $H_R(\bm{k})$, $H_D(\bm{k})$)
are straightforwardly
calculated from Eq.(\ref{interface_H}),
with the corresponding 2D SOC parameters changing sign as well:
\begin{eqnarray}
&& \alpha_{R;v1,v2} = 2 [s_R \mp |\chi_R|\cos(\phi_R - \phi_V)]\,R^2
|\varphi^{\prime}(0)|^2 \qquad
\nonumber\\
&& \beta_{D;v1,v2} = 2 [s_D \mp |\chi_D|\cos(\phi_D - \phi_V)]\,R^2
|\varphi^{\prime}(0)|^2 . \qquad
\label{SOC-consts}
\end{eqnarray}
The scaling of the spin-orbit terms with the electric field $F_z$ is linear, as is the valley splitting, $E_{\rm VS}$.
Here, we have introduced the phases $\phi_R$ and $\phi_D$ for the
Rashba and Dresselhaus terms and $\varphi^{\prime}(0)$ is the
derivative of the $z$-component of an eigenstate of the bulk Hamiltonian ${\cal H}_0$.
These results are similar to the   
strong field limit results
of Ref. \cite{Nestoklon2008}.

Explicit calculation of the $g$-factor change,
based on the interface Hamiltonian Eq.(\ref{interface_H}) and the fact
that an in-plane magnetic field mixes the perpendicular and in-plane motion,
shows that
the in-plane $g$-factor renormalization $\delta g$  is proportional to $\alpha_{R}$, $\beta_{D}$.
For the magnetic field parallel to the $[110]$ direction (as in the experiment)
one obtains the expression:
\begin{equation}
\delta g_{v1,v2} = - \frac{(\alpha_{R;v1,v2} - \beta_{D;v1,v2}) |e|}{\hbar \mu_{\rm B}} \langle z\rangle
\label{g-factor-change}
\end{equation}
where $\mu_{\rm B}$ is the Bohr magneton and $\langle z\rangle \simeq  1.5587\, l_z$
is an average of the $z$-motion in the lowest subband, see Eq.(\ref{bulk_H}).
The $g$-factor scales as $F_z^{2/3}$, which is close to a linear scaling over the range ($\sim 10\%$) of
the experimentally applied electric fields, see Fig. 3b.

We therefore expect from Eq.(\ref{g-factor-change}) that the renormalization $\delta g$ will be of
opposing sign for the two valleys,
following the sign change of the SOC parameters in Eq.(\ref{SOC-consts}).
In particular, the change will be exactly opposite for zero
intra-valley spin-flip
coupling, $s_R, s_D = 0$:
\begin{equation}
\delta g_{v1} = - \delta g_{v2}
\label{opposite_change} .
\end{equation}
Relatively smaller corrections due to non-zero intra-valley spin
flipping,
$s_R, s_D \neq 0$
will generally violate
Eq.~(\ref{opposite_change}), leaving the $g$-factor changes opposite
in sign, but with different absolute value,
$|\delta g_{v1}| \neq |\delta g_{v2}|$.

\begin{figure} [t!]
    \centering
        \includegraphics[width=0.47\textwidth]{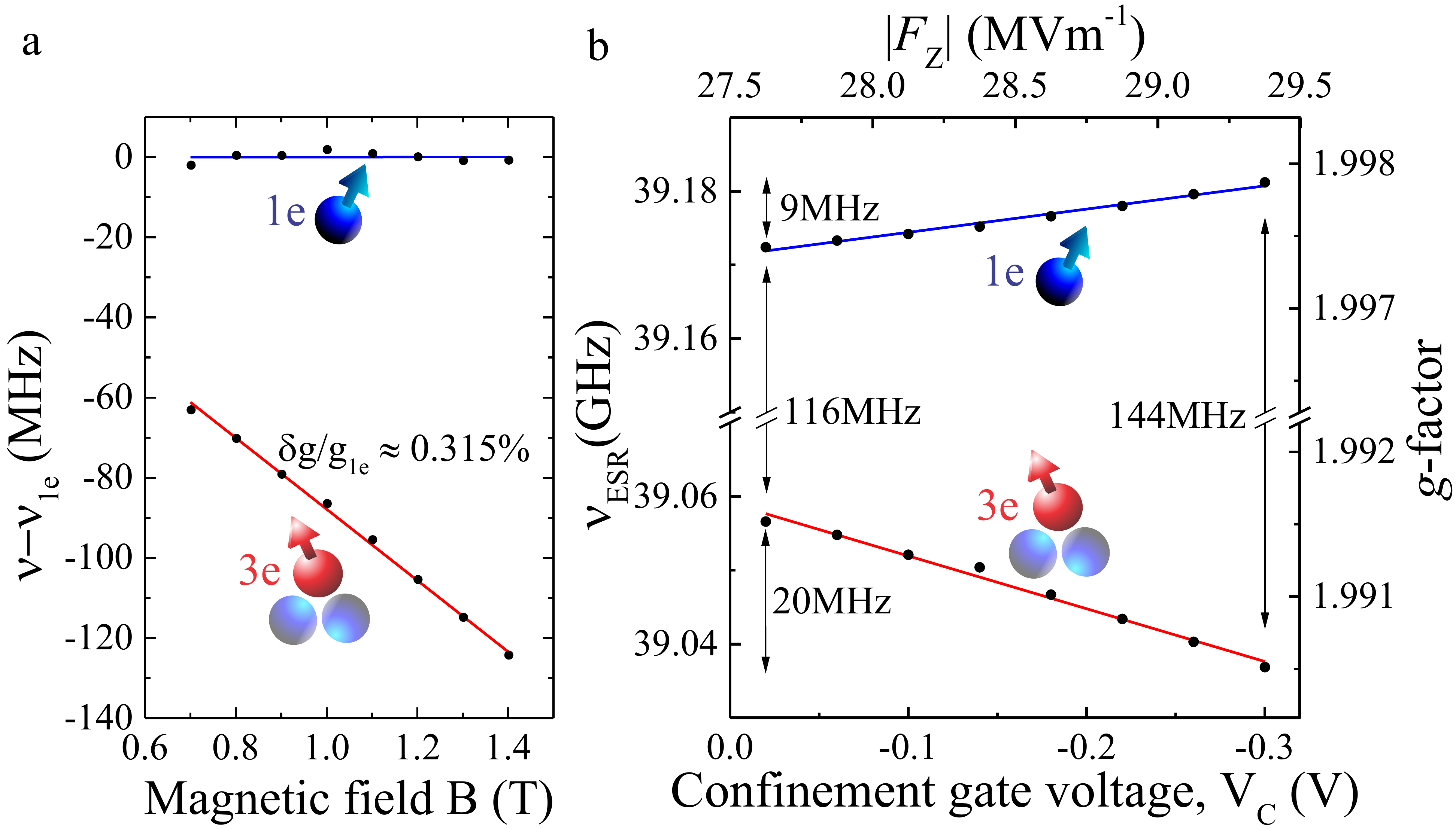}
        \caption{(a) Magnetic field dependence of the resonance frequency of the one- and three-electron qubit, with $F_z$ = 28.25 MV/m.
        The experimental data of both qubit systems has been subtracted by the $g_{1e}$-factor for comparison and
        we have calibrated the dc magnetic field using the crossing point of the one- and three-electron qubit resonance frequencies.
        We find $g_{1e}$ = 1.9975 and $g_{3e}$ = 1.9912. (b) Gate tuned electric field control over the valley $g$-factor at $B_0$ = 1.4015T.}
        \label{fig:3}
\end{figure}

To observe this experimentally, we control the quantum dot electric
field via the plunger gate $PG$ and the confinement gate
$CG$, see Fig. \ref{fig:1}. In Fig. \ref{fig:3}a we show
the magnetic field dependence and in Fig. \ref{fig:3}b we show electrical control over the qubit resonance frequency $\nu_{ESR}$.
The opposite electric field dependence of the $g$-factor for the two valleys is in qualitative agreement
with the prediction of Eq.(\ref{opposite_change}).
Since the resonance frequency of the one-electron qubit
increases with the electric field, while the resonance frequency of the three-electron qubit decreases
(see Fig. \ref{fig:3}b), we
infer from Eq.(\ref{g-factor-change}) that
the  Rashba and Dresselhaus contributions are in this experiment subject to the constraints:
$\delta\chi_{\rm inter-val} \equiv |\chi_R|\cos(\phi_R - \phi_V) - |\chi_D|\cos(\phi_D - \phi_V) > 0$,
$\delta s_{\rm intra-val} \equiv s_R - s_D < 0$.
The change in sign of $\delta g$ is evidence that
the inter-valley spin-flip contributions dominate the intra-valley spin-flip processes and
from the $\delta g$-dependence we estimate
the ratio $\delta\chi_{\rm inter-val}/ |\delta s_{\rm intra-val}| \approx 2.6$.
This observation is consistent with tight-binding
calculations on SiGe quantum wells \cite{Nestoklon2008},
which predict that the inter-valley transitions can be about
an order of magnitude larger than the intra-valley transitions.
Interestingly, the values and signs of the SiGe parameters, as substituted
in Eqs.(\ref{SOC-consts}) and (\ref{g-factor-change}), reproduce also the
correct
qualitative behavior of $\delta g_{v1}$ ($\delta g_{v2}$) that increases (decreases)
with the applied electric field $F_z$, as is observed experimentally in Fig. \ref{fig:3}b.
However, the experimental ratio of the $g$-factor changes is
$|\delta g_{v2}| / |\delta g_{v1}| \simeq 2.2$, 
while that calculated with the Si/SiGe parameters
is $\sim 1$.
%
%
Such differences can be expected due to the much greater band-edge offset in $\rm Si/SiO_2$,
disorder \cite{Friesen2010},
and built-in electric fields which may also
influence the theoretical results.

\begin{figure} [t!]
    \centering
        \includegraphics[width=0.47\textwidth]{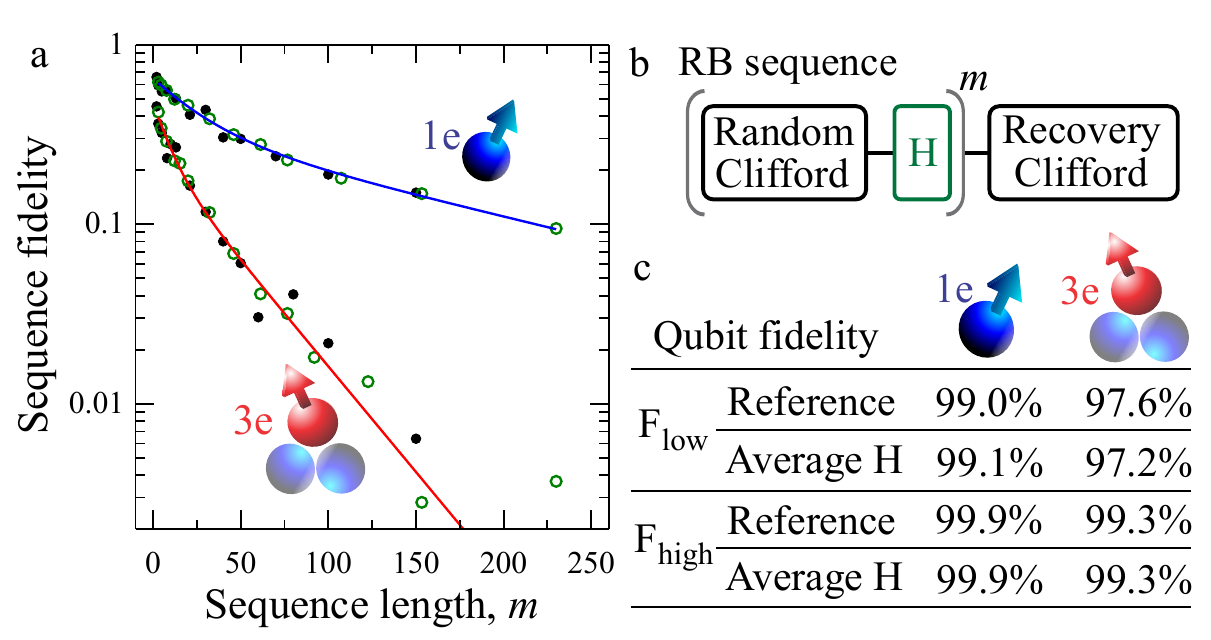}
        \caption{ Clifford based randomized benchmarking (a) Sequence fidelity as a function of the sequence length and
        (b) schematic representation of randomized benchmarking, where $H$ is an interleaved test gate.
        In (a), the black filled circles correspond to standard randomized benchmarking and the green
        open circles to the average of the single qubit gates $[I, \pm X,\pm \frac{1}{2}X,\pm Y,\pm \frac{1}{2}Y]$,
        obtained by interleaved randomized benchmarking and normalizing the sequence length with 2.875/1.875,
        such that it matches the average Clifford  
        gate
         length. Both data sets are fitted
        with a two-fidelity model (see text) and the results are shown in (c), where the standard error
        is smaller than the corresponding gate error.}
        \label{fig:4}
\end{figure}

In order to explore the qubit performance for quantum computation,
we have performed (interleaved) randomized benchmarking (RB)  
\cite{Knill2008, Magesan2012} on the
one-electron qubit \cite{Veldhorst2014} and three-electron qubit, and
all results are shown in Fig. \ref{fig:4}.
In order to eliminate the free fitting parameter $B$, which is a constant
offset parameter present in standard RB  
fits,
we plot the sequence fidelity combination $F=F_{\uparrow} + F_{\downarrow} - 1$,
which approaches zero for infinite sequence length when the assumptions of
RB hold \cite{Fogarty2015}.
When the noise is gate independent, an
exponential decay is expected for both standard and interleaved
randomized benchmarking. However, when low-frequency noise is present, non-exponential decays arise \cite{Fogarty2015}.
This non-exponential decay is due to slow drifts in the resonance frequency,
such that the time ensemble is averaged over sequences with small detuning
(resulting in a high fidelity, $F_{high}$, and a slow exponential decay)
and large detuning
(resulting in a low fidelity, $F_{low}$, and a fast exponential decay).

When such low-frequency noise is present, the fidelity varies over time and we use a
two-fidelity model to analyse the data \cite{Fogarty2015}.
We have fitted the data using $F=A(p^m+q^m)$, where $A$ quantifies the SPAM error and $p$ and $q$ are two polarization parameters,
and in Fig. \ref{fig:4}c we show the corresponding fidelities. The three-electron qubit
has a relatively low fidelity when the noise causes a large detuning ($F_{low}$ $\approx$ 97\%).
 However, when the microwave driving frequency is on resonance, both qubits have a fidelity above certain thresholds for fault tolerant
quantum computing \cite{Fowler2012};
the average single-gate fidelity being $F_{high}$ = 99.9\%
for the one-electron qubit system and $F_{high}$ = 99.3\% for the
three-electron qubit system.
While the three-electron qubit initially shows a non-exponential
decay, for higher $m$ the decay approaches a pure exponential,
indicating that low frequency noise has little impact in this range.
Since both qubit systems are operated with the same setup, we expect
similar calibration errors. This exponential decay is therefore likely
due to high-frequency noise.

The faster decay of the sequence fidelity of the three-electron vs. one-electron qubit
is consistent with a larger sensitivity to electrical noise,
%
as revealed by the larger frequency shift with gate voltage,
$|\delta \nu_{3e} / \delta V| \approx 2.2 |\delta \nu_{1e} / \delta V|$, shown in Fig.\ref{fig:3}b.
The frequency detuning caused by electrical noise results in rotations around the $z$-axis of the
qubit Bloch sphere and opposite to the Rabi driving axis.
By taking the small-angle approximation, we find
that the gate error increases with the square of the noise term. This results
in an error rate that is around 5 times larger for the three-electron qubit, comparable with the
difference in fidelities between the one-electron and three-electron qubits.
It is therefore likely that both qubits are ultimately limited by high-frequency electrical noise.

The recent realizations of single- and two-qubit gates using
isotopically purified silicon quantum dots
\cite{Veldhorst2014, Veldhorst20142} are now revealing the early promises
of silicon as a platform for quantum computation and the possibility of qubit operation with
either one-electron or three-electrons allows more flexibility in scaling these systems. The ultra-narrow
spin-resonance linewidth associated with the long coherence times of these qubits
has pushed silicon into a new regime, where the weak
spin-orbit coupling in silicon becomes not only visible, but also forms
a new tool to control the spin states, as shown here. The
remarkably large electric field control in SiMOS quantum dots provides
further motivation to explore spin-orbit coupling in silicon for qubit
control and spin manipulation.

\begin{acknowledgments}
\textbf{Acknowledgments:}
The authors thank L.M.K. Vandersypen for insightful discussions.
The authors acknowledge support from the Australian Research Council (CE110001027),
the US Army Research Office (W911NF-13-1-0024) and the NSW Node of the Australian National Fabrication Facility.
M.V. acknowledges support from the Netherlands Organization for 
Scientific Research (NWO) through a Rubicon Grant.
The work at Keio has been supported in part by the Grant-in-Aid for Scientific Research by MEXT,
in part by NanoQuine, in part by FIRST, and in part by JSPS Core-to-Core Program.
\end{acknowledgments}

\end{document}